# Cross-sections for nuclide production in $^{56}$Fe target irradiated by 300, 500,750, 1000, 1500, and 2600 MeV protons compared with data on hydrogen target irradiation by 300, 500, 750, 1000, and 1500 MeV/nucleon $^{56}$Fe ions

Yu. E. Titarenko[1], V. F. Batyaev[1], A. Yu. Titarenko[1], M. A. Butko[1], K. V. Pavlov[1],
S. N. Florya[1], R. S. Tikhonov[1], S. G. Mashnik[2], A. V. Ignatyuk[3], N. N. Titarenko[3],
W. Gudowski[4], M. Těšínský[4], C.-M. L. Persson[4],
H. Ait Abderrahim[5], H. Kumawat[6], H. Duarte[7]

[1] Institute for Theoretical and Experimental Physics (ITEP), 117218 Moscow, Russia,
Yury.Titarenko@itep.ru
[2] Los Alamos National Laboratory, Los Alamos, NM 87545, USA
[3] Institute for Physics and Power Engineering, 249020 Obninsk, Russia
[4] Royal Institute of Technology, S - 106 91 Stockholm, Sweden
[5] Centre d'étude de l'énergie nucléaire, 200, Boeretang, Mol, B-2400, Belgium
[6] Bhabha Atomic Research Center, Mumbai –400085, India
[7] CEA/DIF, BP 12 91680 Bruyères-le-Châtel, France

**Abstract.** This work presents the cross-sections for radioactive nuclide production in $^{56}$Fe(p,x) reactions determined in six experiments using 300, 500, 750, 1000, 1500, and 2600 MeV protons of the external beam from the ITEP U-10 proton accelerator. In total, 221 independent and cumulative yields of radioactive residuals of half-lives from 6.6 min to 312 days have been obtained. The radioactive product nuclide yields were determined by direct γ-spectrometry. The measured data have been compared with the experimental data obtained elsewhere by the direct and inverse kinematics methods and with calculation results of 15 different codes that simulated hadron-nucleus interactions: MCNPX (INCL, CEM2k, BERTINI, ISABEL), LAHET (BERTINI, ISABEL), CEM03 (.01, .G1, .S1), LAQGSM03 (.01, .G1, .S1), CASCADE-2004, LAHETO, and BRIEFF. Most of the data obtained here are in a good agreement with the inverse kinematics results and disprove the results of some earlier activation measurements that were quite different from the inverse kinematics measurements. The most significant calculation-to-experiment differences are observed in the yields of the A<30 light nuclei, indicating that further improvements in nuclear reaction models are needed, and pointing out as well to a necessity of more complete experimental measurements of such reaction products.

I. INTRODUCTION

Spallation reactions have been subject a permanent interest during more than 40 years due to production with them a wide variety of residual nuclei, on the one hand, and application of them as intensive neutron sources for both the direct physical and technological researches, and the power accelerator-driven reactor systems, on the other hand [1]. Safety conditions connected with such neutron sources require a sufficiently accurate estimation of the production yields for a large amount of radioactive nuclides accumulated in the corresponding targets and surrounding structural materials. Natural iron is one of the most frequently used structural materials and its nuclear data can be especially important for designing nuclear equipment working under intensive irradiation.

The present-day EXFOR database contains 40 original works [2-41], which (except for [31]) present mainly the cumulative production cross sections of the proton-induced reaction products in natural iron ($^{54}$Fe-5.845%; $^{56}$Fe-91.754%; $^{57}$Fe-2.119%; $^{58}$Fe-0.282%). All the data have been obtained using the proton irradiation of thin iron targets ($p \rightarrow {}^{nat}_{26}Fe$) and the gamma-ray spectrometric analysis to identify the resulting reaction products [2-41].

Recently the precise measurements of the residual production yields for $^{56}$Fe were performed by the inverse kinematics method [42-44]. In case of [43-44], the accelerated $^{56}$Fe ions irradiated a hydrogen target (${}^{56}_{26}Fe \rightarrow {}^{1}_{1}H$) and the yields were measured for the ion kinetic energies of 300, 500, 750, 1000, and 1500 MeV/A using the fragment separator at GSI (Darmstadt).

Comparison of data obtained by different techniques, but at the same energies, is of special interest, because such data are extensively used to verify basic theoretical models included in various high-energy transport codes.

## II. TECHNIQUES FOR EXPERIMENTAL DETERMINATION OF THE RESIDUAL PRODUCTION CROSS SECTIONS

A comprehensive description of the method used for the measurements of the yields or the production cross sections of radioactive products from the proton-induced reactions can be found in works [45-49]. We will briefly discuss below some details of our technique important for the present experiment.

### A. Determination of the residual production rates

For a consistent estimation of uncertainties of measured cross sections the concept of independent and cumulative reaction rates [48, 49] is usually introduced, which can be determined by the following relations:

$$R^{ind} = \sigma^{ind}(E) \cdot \Phi(E) \quad \text{and} \quad R^{cum} = \sigma^{cum}(E) \cdot \Phi(E), \qquad (1)$$

where $\sigma^{ind}(E)$ and $\sigma^{cum}(E)$ are, respectively, the independent and cumulative cross sections of a nuclide production and $\Phi(E)$ is the proton flux density.

Accumulation of the reaction products during the proton beam irradiation and afterwards is described by the set of kinetic equations, the solution of which for the reaction rates depends from both the decay constants of the involved radioactive nuclei and the irradiation conditions. Analytical expressions for the reaction rates have been obtained for the case of two- and three-link decay chains. In the present experiment the yield of nuclides produced at the two-link decay chains only were measured and the corresponding expressions for the reaction rates can be written in the form

$$R_1^{cum/ind} = \frac{A_1}{N_T \cdot \eta_1 \cdot \varepsilon_1} \cdot \frac{1}{F_1} \qquad (2).$$

$$R_1^{cum/ind} = \frac{A_2'}{N_T \cdot \eta_2 \cdot \varepsilon_2 \cdot v_{12}} \cdot \frac{\lambda_2 - \lambda_1}{\lambda_2} \cdot \frac{1}{F_1}, \qquad (3),$$

$$R_2^{ind} = \left( \frac{A_2''}{F_2} + \frac{A_2'}{F_1} \cdot \frac{\lambda_1}{\lambda_2} \right) \cdot \frac{1}{N_T \cdot \eta_2 \cdot \varepsilon_2}, \qquad (4),$$

$$R_2^{cum} = R_2^{ind} + v_{12} \cdot R_1^{cum/ind} = \left( \frac{A_2'}{F_1} + \frac{A_2''}{F_2} \right) \cdot \frac{1}{N_T \cdot \eta_2 \cdot \varepsilon_2}, \qquad (5).$$

where $A_1$, $A_2'$, and $A_2''$ are the parameters determined by the least square fitting of the experimental decay curves of the nuclei, with subscripts *1* and *2* designating the mother and daughter nuclides, respectively; $N_T$ is the number of nuclei in an irradiated sample; $\eta_1$ and $\eta_2$ are the γ-ray abundances; $\lambda_1$ and $\lambda_2$ are the decay constants; $\varepsilon_1$ and $\varepsilon_2$ are the absolute spectrometer effectivenesses at γ-energies $E_1$ and $E_2$; $v$ is the branching factor, i.e. the probability for a mother nuclide to turn unto its daughter; $F_1$ and $F_2$ are functions to be calculated as $F_i = \left(1 - e^{-\lambda_i t_{irr}}\right)$; $t_{irr}$ is the irradiation time.

**B. Determination of the mean proton flux density**

The mean proton flux density $\Phi(E)$ was determined using the $^{27}Al(p,x)^{22}Na$ monitor reaction. Fig. 1 compiles the experimental data of 30 works made from 1955 to 1997 in 25 MeV -3.0 GeV energy range. Out of the data presented, the data of but two works by Tobailem, 1981 ($E_p$>200 MeV) and Steyn, 1990 ($E_p$<200 MeV) were chosen for monitoring purposes. In the case of energies that were not supported with experimental cross sections, they were calculated via linear interpolation of the cross section logarithms at the boundaries of the given range.

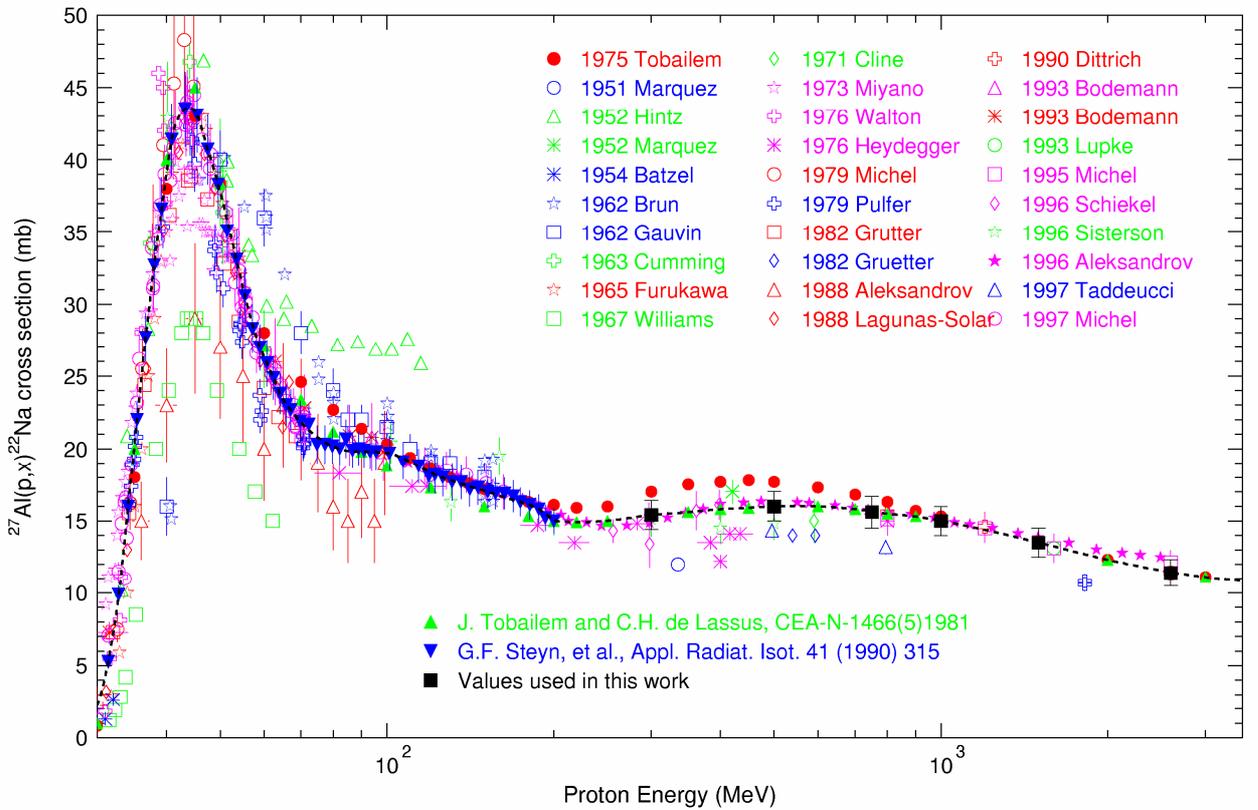

FIG. 1. The $^{27}Al(p,x)^{22}Na$ reaction excitation function in the 25-3000 MeV proton energy range

In this case, the time- and sample area-averaged density of proton flux and its error are calculated as

$$\widehat{\Phi} = \frac{R^{22Na}}{\sigma^{22Na}} \qquad \frac{\Delta\widehat{\Phi}}{\widehat{\Phi}} = \sqrt{\left(\frac{\Delta R^{22Na}}{R^{22Na}}\right)^2 + \left(\frac{\Delta \sigma^{22Na}}{\sigma^{22Na}}\right)^2}, \qquad (6)$$

where $R^{22Na}$ is the $^{22}Na$ production rate and $\sigma^{22Na}$ is the monitor-reaction cross section at a given energy. The analytical expression to calculate the $^{22}Na$ production rate is the same as (2).

## C. Preparation of samples

The samples have the diameter of 10.5-mm and were prepared by pressing the $^{56}$Fe-enriched fine-dispersed iron powder ($^{54}$Fe – 0.3%, $^{56}$Fe – 99.5±0.1%, $^{57}$Fe – 0.2%, $^{58}$Fe <0.05%). The Al monitors were cut off from a foil and pressed then into the same master form to provide identity of the parameters. The samples and the monitors were weighed. After that, each of the sample-Al interlayer-Al monitor sandwiches was sealed into a polyethylene package to preserve the sandwich geometry and was then directed to irradiation.

The impurities amounted to ~ 0.091% in the iron powder and to ~ 0.01% in the Al foil.

## D. Irradiation of samples

The experimental samples were irradiated using the U-10 ITEP proton synchrotron. The synchrotron serves as a ring facility with the 25 MeV-proton energy injector and the highest accelerated proton energy of 9.3 GeV. The accelerated proton beam with a given energy (ranging from 40 to 9300 MeV) consists from 4 bunches of a ~250 ns duration each and is directed from the synchrotron ring to the transport channel that provides proton extraction of a ~2·10$^{11}$ proton/pulse intensity, an elliptic cross section with 10x15 mm axes, ~1 µs total duration, and ~15 pulse/min pulse repetition rate. The transport channel and its elements are described in detail in [49].

During irradiation runs, a polyethylene package with a sample sandwich therein was fixed with a scotch to the center of a 50x50 mm, 0.1 mm thick Al plate placed in a task-oriented holder perpendicularly to the proton beam. The selected geometry precludes $^{24}$Na, $^{22}$Na, and $^{7}$Be produced at the Al monitor from accumulating in the sample. The monitor provides the total control of a proton beam intensity. Table 1 presents the parameters of the $^{56}$Fe sample irradiation.

TABLE 1. The $^{56}$Fe target irradiation parameters.

| Energy (GeV) | Mass of sample (g) | Mass of a monitor (g) | Irradiation time (min) | Mean proton flux (p/(cm$^2$ s) x 10$^{10}$) |
|---|---|---|---|---|
| 0.3 | 0.2422 | 0.0480 | 35 | 5.23±0.38 |
| 0.5 | 0.2447 | 0.0486 | 33 | 6.73±0.48 |
| 0.75 | 0.2433 | 0.0499 | 30 | 7.45±0.64 |
| 1.0 | 0.2429 | 0.0494 | 35 | 6.11±0.46 |
| 1.5 | 0.2429 | 0.0485 | 36 | 4.44±0.37 |
| 2.6 | 0.2000 | 0.1202 | 30 | 1.70±0.15 |

During the experiment, the beam parameters were on-line controlled using current transformer together with the task-oriented digital-control PC plate with a 2 ns time resolution. The digitized data were recorded to a 2 μs full sweep file (~1000 values) used to calculate the amplitudes of each proton pulse. Such data are needed to allow in detail for the decay of nuclei under irradiation, which is especially important for the short-lived nuclides.

**E. Proton beam energy**

The proton energy on the target $(E_{sample})$ is determined to be a difference between the kinetic energy of protons extracted from an accelerator ring $(E_0)$ and the loss in transport channel membrane, in air gap, in polyethylene, and in experimental sample half width $(E_{loss})$

$$E_{sample} = E_0 - E_{loss} \qquad (7)$$

The kinetic energy of the protons both circulating and extracted to the transport channel is determined as

$$E_0 = \frac{m_p \cdot c}{\sqrt{c^2 - L^2 \cdot f_r^2}} - m_p \qquad (8)$$

where $E_0$ is the kinetic energy of a circulating proton; $m_p$= 938.26 is proton mass; $L$=251.21 m is the closed orbit length; $c$=2.99776 $10^8$ m/s is speed of light.

The $f_r$ value is multiple to the accelerating radio frequency:

$$f_a = h \cdot f_r \qquad (9)$$

where $h$=4 is he number of harmonics; $f_a$ is the accelerating radio frequency.

The beam energy loss $(E_{loss})$ due to passage through the channel structure elements and in the target proper were calculated as $\delta E=(dE/dx)\cdot X$, which formula can be considered sufficiently correct because the thickness *(X)* traversed by the beam is insignificant, so the specific loss for ionization *dE/dx* may well be taken constant.

**F. Measurement and processing of the gamma-ray spectra**

The reaction products in the irradiated samples and monitors were identified by measuring the typical discreet gamma-ray spectra of produced nuclei. For that purpose, the gamma-ray spectrometer based on a coaxial Ge-detector was used with a 1.8 keV energy resolution for the 1332 keV gamma-line of $^{60}$Co. The admissible spectrometer parameters and measurement modes were determined in preliminary experiments, and they were strictly controlled under subsequent measurements. To reduce the spectrometer load during the initial gamma-spectrum measurements, an experimental sample irradiated was placed at a considerable height over the detector surface.

The temperature stability, the cascade summation effects, the maximum spectrometer load, and the absolute height-energy effectiveness *ε(E,H)* of gamma-ray detection were controlled. The technique for determining *ε(E,H)* has been described in detail in Ref. [49], which presents the analytical expressions to calculate the latter for the gamma-ray energies from 90 keV to 2600 keV and the heights between 40 and 1240 mm.

The gamma-ray spectra were analyzed with the GENIE-2000 code. After automated packet processing the measured spectra, the code permits operating for each spectrum with the interactive fitting mode to additionally examine the results of tentative processing.

The processed spectra were merged into a single file to form an input file for the SIGMA code. The code plots the time variations in the selected gamma-line intensities and identifies the corresponding nuclides in accordance the NUDAT database and the additional gamma-ray transition schemes presented in Refs. [50-52]. For identified nuclides the code calculates finally the nuclide production rate in accordance with the above formulas (2)-(5). Fig. 2 shows some examples of the measured count rates and fitted decay curves.

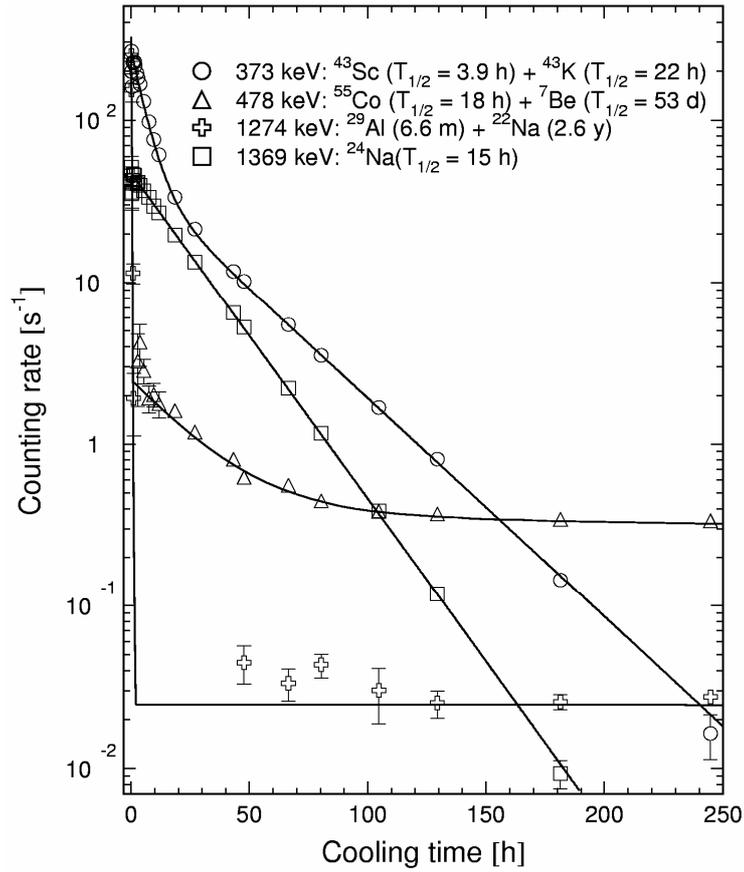

FIG. 2. Examples of the measured count rates and fitted decay curves.

Table 2 presents the nuclear physics characteristics of nuclides produced in proton-irradiated $^{56}$Fe used to identify the nuclides and to determine their cross sections.

TABLE 2. Nuclear-physics characteristics of nuclides.

| Nuclide | Half-life | Gamma-energies ($E_\gamma$, keV) and abundances ($Y_\gamma$, %) |
|---|---|---|
| $^{57}$Co | 271.74d | *122.1 (85.60), 136.5 (10.68)* |
| $^{56}$Co | 77.233d | *846.8 (99.94), 1037.8 (14.17), 1238.3 (66.9), 1360.2 (4.29), 1771.3 (15.47), 2015.2 (3.04), 2034.8 (7.89), 2598.5 (17.3)* |
| $^{55}$Co | 17.53h | *477.2 (20.2), 931.1 (75.0), 1408.5 (16.9)* |
| $^{53}$Fe | 8.51m | *377.9 (42.0)* |
| $^{52}$Fe | 8.275h | *168.7 (99.0)* |
| $^{56}$Mn | 2.5789h | *846.7 (98.), 1810.7 (27.2), 2113.1 (14.3)* |
| $^{54}$Mn | 312.11d | *834.8 (99.9760)* |
| $^{52m}$Mn | 21.1m | *1434.1 (99.8), 1727.5 (0.220)* |
| $^{52}$Mn | 5.591d | *744.2 (90.0), 848.2 (3.32), 935.5 (94.5), 1246.3 (4.21), 1333.7 (5.07), 1434.1 (100.0)* |
| $^{51}$Cr | 27.7025d | *320.1 (9.92)* |
| $^{49}$Cr | 42.3m | *90.6 (53.2), 152.9 (30.3)* |
| $^{48}$Cr | 21.56h | *112.3 (96.0), 308.2 (100.0)* |
| $^{48}$V | 15.9735d | *928.3 (0.77), 944.1(7.76), 983.5 (100.0), 1312.1 (97.5)* |
| $^{48}$Sc | 43.67h | *175.4 (7.48), 983.5 (100.1), 1037.5 (97.6), 1312.1 (100.1)* |
| ⁴⁷Sc | 3.3492d | *159.4 (68.3)* |
| ⁴⁶Sc | 83.79d | *889.3 (99.9840), 1120.5 (99.9870)* |
| ⁴⁴ᵐSc | 58.61h | *270.9 (86.7), 1001.8 (1.20), 1126.1 (1.20)* |
| ⁴⁴Sc | 3.97h | *1499.5 (0.908), 2656.5 (0.112)* |
| ⁴³Sc | 3.891h | *372.8 (22.5)* |
| ⁴⁷Ca | 4.536d | *1297.1 (71.0)* |
| ⁴⁴K | 22.13m | *1499.4 (7.80), 2150.8 (23.0)* |
| ⁴³K | 22.3h | *372.8 (86.80), 396.9 (11.85), 593.4 (11.26), 617.5 (79.2)* |
| ⁴²K | 12.360h | *1525.0 (18.08)* |
| ⁴¹Ar | 109.34m | *1294 (99.10)* |
| ³⁹Cl | 55.6m | *250.3 (46.3), 1267.0 (53.6), 1518.0 (39.2)* |
| ³⁸Cl | 37.24m | *1643.0 (31.9), 2167.0 (42.4)* |
| ³⁴ᵐCl | 32.00m | *146.4 (40.5), 1177.0 (14.09), 2127.0 (42.8)* |
| ³⁸S | 170.3m | *1942.0 (83.0)* |
| ²⁹Al | 6.56m | *1273.3 (90.6)* |
| ²⁸Mg | 20.915h | *400.7 (36.6), 941.4 (38.3), 1342.3 (52.6)* |
| ²⁷Mg | 9.458m | *843.8 (71.8), 1014.4 (28.0)* |
| ²⁴Na | 14.9590h | *1369.0 (100.0)* |
| ²²Na | 2.6019y | *1274.5 (99.944)* |
| ⁷Be | 53.29d | *477.6 (10.52)* |

### III. EXPERIMENTAL RESULTS AND THEIR UNCERTAINTIES

Using formula (1), the production cross sections of independent and cumulative ⁵⁶Fe(p,x) reaction products and their uncertainties can be calculated to be

$$\sigma^{ind/cum} = \frac{R^{ind/cum}}{\widehat{\Phi}} \qquad \frac{\Delta\sigma^{ind/cum}}{\sigma^{ind/cum}} = \sqrt{\left(\frac{\Delta R^{ind/cum}}{R^{ind/cum}}\right)^2 + \left(\frac{\Delta\widehat{\Phi}}{\widehat{\Phi}}\right)^2}, \qquad (10)$$

where $R^{ind/cum}$ is the ⁵⁶Fe(p,x) production rate, calculated via formulas (2-5); $\widehat{\Phi}$ is the mean proton flux density calculated by (6).

From our experiments, we have obtained 221 values of the residual production cross sections (yields) at various energies of incident protons, which include 54 independent yields (i), 18 independent yield for isomeric states of residual nuclides (i(m)), 17 sums of independent ground and isomeric states (i(m+g)), 132 cumulative and supracumulative yields (c, c*). The supracumulative yield concept is discussed in Ref. [46]. Numerical values of obtained data are presented in Table 3 and plotted in Figs. 3-5 [1]. For comparison, the figures present also the plots of other available data, as well as the results of cross section calculations with various codes discussed below.

As seen from Table 2, a few gamma-lines were used instead a single one to estimate the reaction rates for more than a half of the measured cross sections. Therefore, a special method was developed to calculate the averaged rates. The respective analytical expressions are presented in [49].

The above techniques is featured in that the initial averaging step uses the relative yields *m* of different gamma-lines and the relative spectrometer effectiveness at the energies of given gamma-lines, thereby permitting, according to formulas (2)-(5), the mean relative reaction rate to be calculated correctly. During the second step, the relative-to-absolute gamma-line yield transition coefficient is used together with the absolute activity of standard sources to determine the mean absolute production rate of each nuclide.

TABLE 3. Experimental values of $^{56}$Fe(p,x) product cross sections for $E_p$ = 300, 500, 750, 1000, 1500 and 2600 MeV protons.

| Product | Type | $T_{1/2}$ | Cross section σ±Δσ (mbarn) | | | | | |
|---|---|---|---|---|---|---|---|---|
| | | | $E_p$=300 MeV | $E_p$=500 MeV | $E_p$=750 MeV | $E_p$=1000 MeV | $E_p$=1500 MeV | $E_p$=2600 MeV |
| $^{57}$Co | i | 271.74d | 0.091±0.008 | 0.125±0.010 | 0.189±0.018 | 0.245±0.021 | 0.314±0.029 | 0.369±0.036 |
| $^{56}$Co | i | 77.233d | 1.42±0.12 | 1.02±0.08 | 0.908±0.083 | 0.976±0.080 | 0.960±0.086 | 1.02±0.10 |
| $^{55}$Co | i | 17.53h | 1.00±0.08 | 0.570±0.046 | 0.402±0.041 | 0.375±0.039 | 0.329±0.030 | 0.276±0.027 |
| $^{53}$Fe | c* | 8.51m | 3.28±0.60 | 4.39±0.95 | 2.14±0.33 | 2.93±0.69 | 2.62±0.32 | 2.39±0.40 |
| $^{52}$Fe | c | 8.275h | 0.628±0.050 | 0.471±0.037 | 0.372±0.034 | 0.349±0.028 | 0.301±0.027 | 0.230±0.021 |
| $^{56}$Mn | c | 2.5789h | 0.246±0.021 | 0.442±0.044 | 0.644±0.060 | 0.791±0.066 | 0.852±0.076 | 0.854±0.079 |
| $^{54}$Mn | i | 312.11d | 44.9±3.6 | 42.0±3.3 | 40.4±3.7 | 42.4±3.5 | 39.4±3.5 | 32.7±3.0 |
| $^{52m}$Mn | i(m) | 21.1m | 9.98±0.85 | 9.49±0.83 | 7.71±0.73 | 7.37±0.64 | 6.72±0.62 | 5.33±0.51 |
| $^{52m}$Mn | c | 21.1m | 10.6±0.9 | 9.97±0.87 | 8.17±0.82 | 7.76±0.67 | 7.08±0.69 | 5.55±0.53 |

---

[1] Beside the data presented in Table 3, Figs 3-5 show the $^{56}$Fe(p,x) reaction yields for 250, 400, 600, 800, 1200, and 1600 MeV proton energies as measured at ITEP under the current ISTC Project#3266. The complete numerical data will be presented in the Final Technical Report on that Project.

| Nuclide | Type | Half-life | | | | | | |
|---|---|---|---|---|---|---|---|---|
| $^{52}$Mn | c | 5.591d | 14.4±1.2 | 12.0±1.0 | 10.0±0.9 | 9.63±0.82 | 8.56±0.78 | 6.80±0.63 |
| $^{51}$Cr | c | 27.7025d | 52.9±4.3 | 47.7±3.8 | 41.9±3.9 | 41.0±3.4 | 35.5±3.2 | 28.3±2.6 |
| $^{49}$Cr | c | 42.3m | 7.08±0.57 | 7.31±0.58 | 6.26±0.58 | 5.92±0.49 | 5.09±0.47 | 3.96±0.37 |
| $^{48}$Cr | c | 21.56h | 0.929±0.080 | 0.973±0.078 | 0.875±0.081 | 0.836±0.070 | 0.690±0.063 | 0.504±0.047 |
| $^{48}$V | c | 15.9735d | 22.0±1.8 | 23.0±1.8 | 21.1±1.9 | 20.6±1.7 | 17.4±1.5 | 13.3±1.2 |
| $^{48}$Sc | i | 43.67h | 0.313±0.039 | 0.473±0.040 | 0.553±0.053 | 0.607±0.060 | 0.547±0.055 | 0.437±0.045 |
| $^{47}$Sc | i | 3.3492d | 2.32±0.19 | 3.26±0.26 | 3.58±0.33 | 3.73±0.31 | 3.40±0.31 | 2.66±0.24 |
| $^{47}$Sc | c | 3.3492d | 2.36±0.19 | 3.30±0.26 | 3.62±0.34 | 3.80±0.32 | 3.44±0.31 | 2.71±0.25 |
| $^{46}$Sc | i(m+g) | 83.79d | 6.93±0.56 | 9.51±0.75 | 10.2±0.9 | 10.6±0.9 | 9.33±0.84 | 7.16±0.67 |
| $^{44m}$Sc | i(m) | 58.61h | 5.58±0.44 | 8.45±0.66 | 9.46±0.86 | 9.87±0.80 | 8.94±0.79 | 6.33±0.57 |
| $^{44}$Sc | i(m+g) | 3.97h | 10.3±1.5 | 15.9±2.0 | 18.8±4.0 | 18.9±2.5 | 17.2±1.8 | 12.8±1.3 |
| $^{44}$Sc | i | 3.97h | 6.05±0.63 | 9.6±1.1 | 11.3±1.9 | 11.1±1.4 | 7.8±1.2 | 6.69±0.75 |
| $^{43}$Sc | c | 3.891h | 3.07±0.27 | 5.14±0.44 | 6.07±0.59 | 6.49±0.58 | 5.66±0.54 | 4.08±0.39 |
| $^{47}$Ca | c | 4.536d | 0.031±0.002 | 0.032±0.002 | 0.045±0.004 | 0.058±0.005 | 0.051±0.005 | 0.053±0.005 |
| $^{44}$K | c* | 22.13m | -- | -- | 0.188±0.080 | 0.254±0.069 | 0.267±0.067 | 0.383±0.365 |
| $^{43}$K | c | 22.3h | 0.467±0.037 | 0.962±0.076 | 1.33±0.12 | 1.53±0.14 | 1.48±0.13 | 1.16±0.11 |
| $^{42}$K | i | 12.360h | 1.62±0.13 | 3.28±0.26 | 4.39±0.41 | 5.08±0.42 | 4.84±0.44 | 3.89±0.35 |
| $^{41}$Ar | c | 109.34m | 0.166±0.014 | 0.416±0.034 | 0.650±0.060 | 0.809±0.067 | 0.846±0.076 | 0.699±0.064 |
| $^{39}$Cl | c | 55.6m | 0.074±0.010 | 0.227±0.022 | 0.387±0.037 | 0.500±0.044 | 0.556±0.053 | 0.519±0.051 |
| $^{38}$Cl | i(m+g) | 37.24m | 0.263±0.047 | -- | 1.44±0.14 | 1.95±0.17 | 2.18±0.20 | 1.67±0.17 |
| $^{38}$Cl | c | 37.24m | 0.276±0.028 | 0.913±0.081 | 1.48±0.14 | 1.99±0.17 | 2.23±0.21 | 1.71±0.17 |
| $^{34m}$Cl | i(m) | 32.00m | 0.086±0.013 | 0.287±0.031 | 0.653±0.063 | 0.903±0.079 | 1.05±0.10 | 0.871±0.084 |
| $^{38}$S | c | 170.3m | -- | -- | 0.029±0.003 | 0.046±0.004 | 0.049±0.004 | 0.046±0.004 |
| $^{29}$Al | c | 6.56m | -- | 0.363±0.041 | 0.84±0.12 | 1.61±0.41 | 2.48±0.26 | 1.59±0.41 |
| $^{28}$Mg | c | 20.915h | 0.008±0.001 | 0.048±0.007 | 0.111±0.011 | 0.209±0.020 | 0.355±0.034 | 0.392±0.036 |
| $^{27}$Mg | c | 9.458m | -- | 0.174±0.022 | 0.463±0.053 | 0.59±0.11 | 1.30±0.13 | 1.54±0.16 |
| $^{24}$Na | c | 14.9590h | 0.072±0.020 | 0.293±0.034 | 0.977±0.097 | 1.80±0.16 | 3.05±0.28 | 3.69±0.34 |
| $^{22}$Na | c | 2.6019y | 0.070±0.014 | 0.264±0.024 | 0.672±0.064 | 1.35±0.11 | 2.35±0.21 | 3.20±0.34 |
| $^{7}$Be | i | 53.29d | 0.891±0.076 | 1.90±0.16 | 3.19±0.30 | 4.87±0.40 | 7.30±0.66 | 8.91±0.82 |

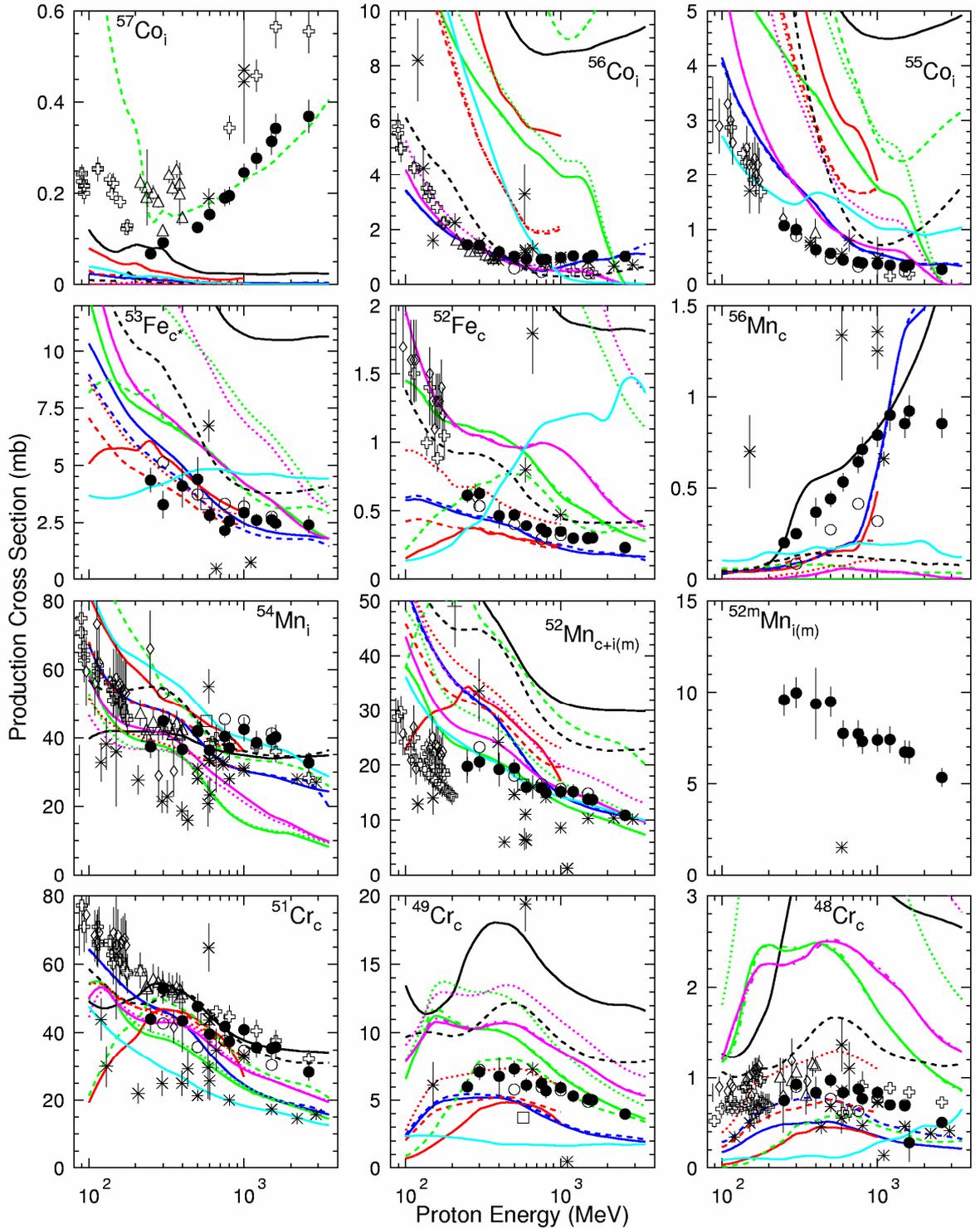

F IG. 3. $^{56}$Fe(p,x) nuclide production cross sections measured at ITEP (●), GSI (○, [43]), R. Michel et al. (⊕,[2-6]), Th. Shiekele et al.(△,[7]), M. Fassbender et al. (◊,[8,9]), W. R. Webber et al. (□,[42]), others (✳,[10-41]), INCL/MCNPX (solid black), BRIEFF1.5.4g (dashed black), CEM03.01 (solid green), CEM2k/MCNPX (dashed green), CEM03.G1 (dotted green), CEM03.S1 (dashed-dotted green), BERTINI (MCNPX − solid blue, LAHET - dashed blue), ISABEL (MCNPX − solid red, LAHET - dashed red, LAHETO – dotted red), LAQGSM03.01 (solid magenta), LAQGSM03.G1 (dotted magenta), LAQGSM03.S1 (dashed-dotted magenta), CASCADE-2004 (cyan).

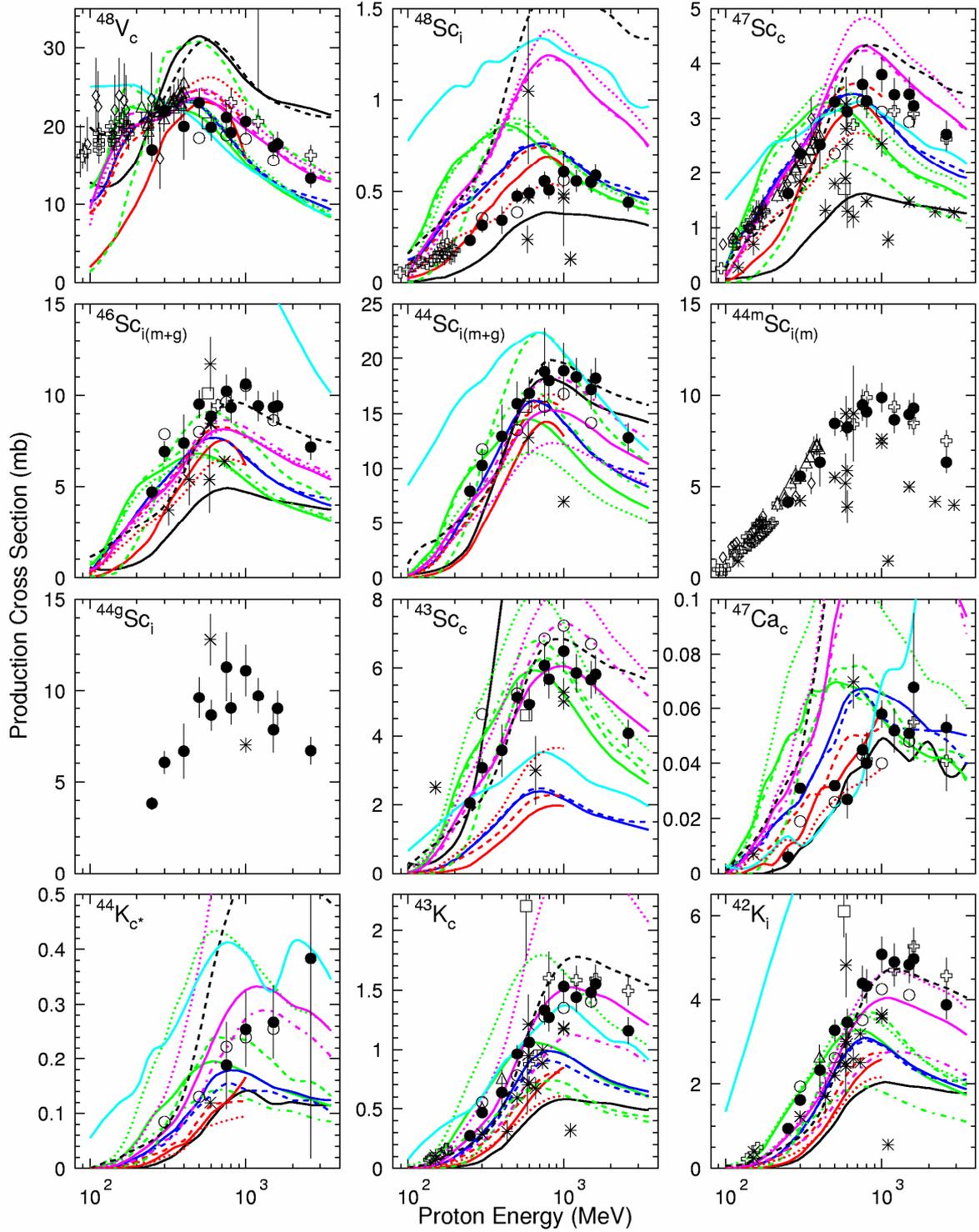

FIG. 4. Continuation of Fig. 3.

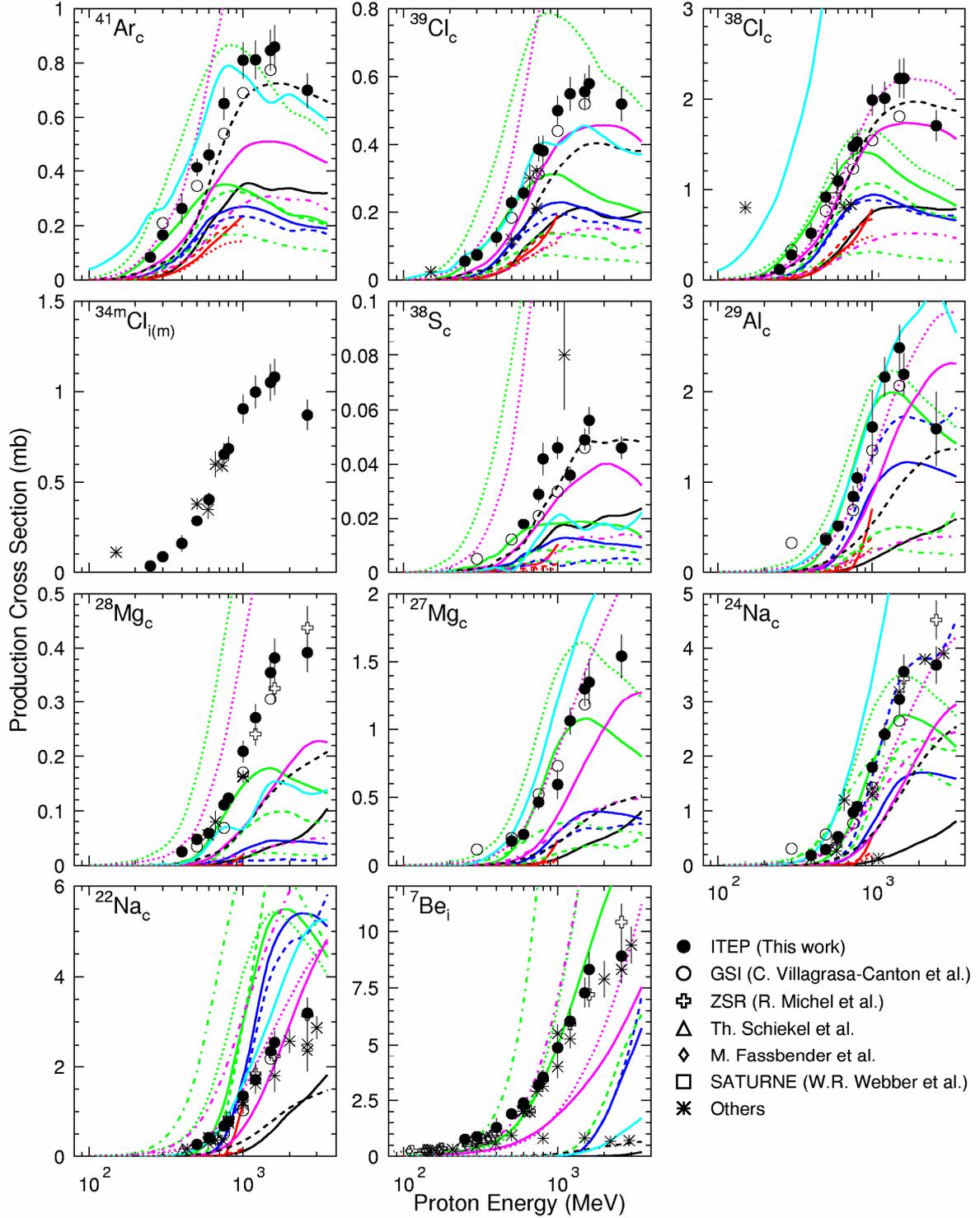

FIG. 5. Continuation of Fig. 4.

## A. Analysis of errors

For an analysis of data obtained by different laboratories and comparison them with theoretical calculations it is necessary to take into consideration both the uncertainties of the measured reaction yields and the uncertainties of the incident proton energies.

The proton energy uncertainties arise from the U-10 synchrotron operations and from the geometric parameters of the transport channel that provides the 40-3000 MeV proton beam extraction. The constancy of proton orbit in a ring and the high accuracy ($\sim 10^{-2}$%) of determining the signal $f_a$, that formula (8) requires in calculating the proton kinetic energy permit the energy error to be determined within ~0.5%.

The uncertainties in the reaction rates of residual nuclides produced in experimental sample and of $^{22}$Na produced in Al monitor are due to the following two main factors. The first arises from the performances of the equipment used, from the γ-spectrometer operations, from the precision degree of the calibration gamma-ray sources, from the analytical balance accuracy, and from the certifying accuracy of the irradiated material compositions. The second factor is due to the uncertainties in the nuclear data from different databases and publications, namely, PCNUDAT [50], Tables of Isotopes (8th Edition) [51], ENSDF [52], Photon cross section from 1 keV to 100 MeV for elements Z=1 to Z=100 [53], and the original works (the $^{27}$Al(p,x)$^{22}$Na monitor reaction cross sections) [54, 55].

As shown by calculations, the accuracy in experimental determining the reaction rates is within 3.1% - 95% range (with 6.2% being the mean) and the accuracy of determining the mean proton flux is 7.1 -8.6%. Therefore, the accuracy of the above presented production cross sections (yields) of radioactive residuals is 7.8 - 95%, with 10.7% being the mean. Fig. 6 shows the distributions of the errors in reaction rates and cross sections.

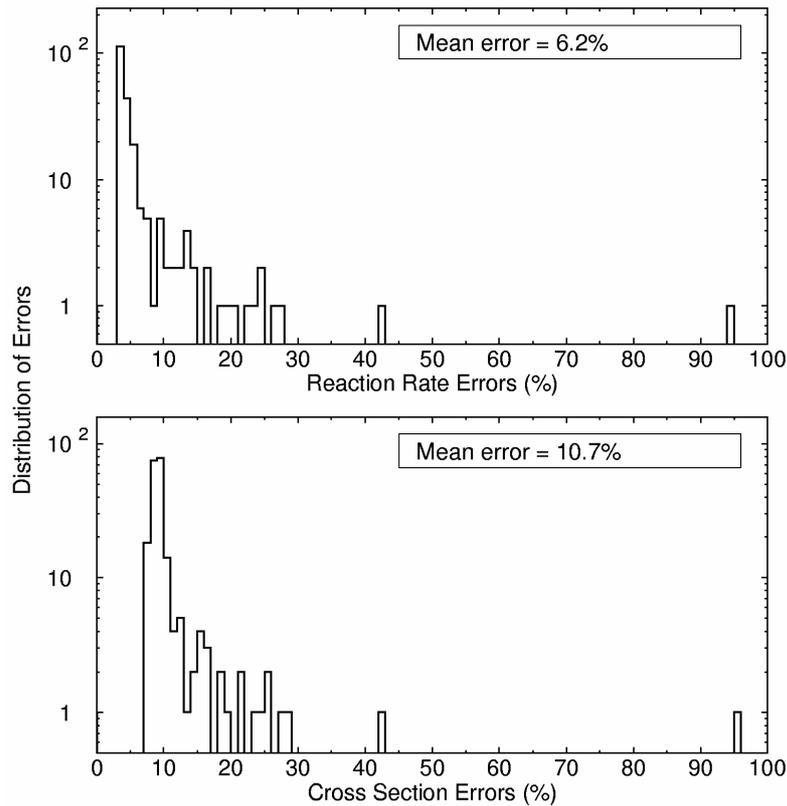

FIG. 6. Distributions of the errors in reaction rates (top plot) and in cross sections (bottom plot).

The main uncertainties of formulas (2-6) contributors are as following:

- The statistical uncertainty of count rate in total absorption peak with due allowance for correction of unresolved γ-lines, of background under the peak, and γ-spectrum transients, as well as for the spectrometer dead time and count loss, to the total error of the nuclide production rate varies from 1.0% to 95%.

- The uncertainty in the correction for γ-absorption in samples varies from 0.03% to 0.07%.

- The uncertainty in the absolute spectrometer detection effectiveness with due allowance for uncertainty in cascade summation effects and in contributions from the peaks of double and single ejections varies from 3.3% to 3.4%.

- The contribution of the uncertainty in the γ-line abundance varies from 0% to 7.1%.

- The contribution of the uncertainty in the number of experimental sample nuclei is about 0.05%.

- The $^{27}$Al(p,x)$^{22}$Na monitor reaction cross section uncertainty to the proton flux density uncertainty varies from 6.3% to 7.9%.

As shown by the above analysis, the main uncertainties are presented by statistical errors, absolute spectrometer detection effectiveness, and γ-abundance uncertainty in the case of experimental samples, and by absolute spectrometer detection effectiveness only in the case of monitors.

The uncertainty in the proton flux density arises from the monitor reaction uncertainty only.

The uncertainties determined by the sample compositions, by the uncertainty in the correction that allows for the measured nuclide production from secondaries, by the decay constant uncertainties, and by the uncertainties in the irradiation, decay, and measurement times ($t_{irr}$, $t_d$, t) were disregarded because of their smallness, which did not actually affect the uncertainties in the results.

**B. Comparison with the data obtained elsewhere**

We compare our current data with previous measurements from 42 works [2-44]. For convenience, we divide all data into seven groups and show them with different symbols in Figs. 3-5, respectively: Our current data presented in Table 3 are shown with filled circles; the GSI inverse kinematics data [43] are shown with open circles; the data by Michel *et al.* [2-6] are shown with open crosses; the data by Webber *et al.* [42] are shown with open squares; the data by Schiekel *et al.* [7] are shown with open triangles; the data by Fassbender *et al.* [8, 9] are shown with open diamonds; the data from the rest of 32 works [10-41] have been united into a single group and are shown with stars.

Using the procedure described below to calculate cumulative yields from independent ones, as well as the mean square deviation factor <F> calculated with formula (13), which is usually employed to analyze various theoretical and experimental data, the current ITEP and GSI [43] data were compared quantitatively. Within this approach, all products were divided into two groups of spallation (A>30) and fragmentation (A<30) products, respectively. Table 6 presents the comparison results.

As the data from the rest of the works [2-44] have been obtained at different energies, we compare them with our results only qualitatively, in Figs. 3-5. Comparing our results with the data by Michel *et al.* [2-6], we can observe some serious discrepancies only for $^{57}$Co at all energies, for $^{48}$Cr at energies above 1 GeV, and for $^{52}$Mn at energies below 200 MeV, while other cross sections agree reasonably well. With the data by Schiekel *et al.* [7], we see some big discrepancies only for $^{57}$Co, and with the data by Fassbender *et al.* [8, 9], only for $^{52}$Mn.

TABLE 4. The mean square deviation factor <F> of the ITEP and GSI data for different incident proton energies and mass numbers of the products groups.

|  | Product mass (A), proton energy ($E_p$, GeV) | | | | | All energies; all products |
|---|---|---|---|---|---|---|
|  | 300 | 500 | 750 | 1000 | 1500 |  |
| ITEP – GSI (A<30) | 3.14 | 1.67 | 1.33 | 1.25 | 1.14 |  |
| ITEP – GSI (A>30) | 1.34 | 1.28 | 1.28 | 1.28 | 1.25 | 1.34 |
| ITEP – GSI (all A) | 1.53 | 1.37 | 1.28 | 1.28 | 1.23 |  |

The discrepancy in the case of $^{57}$Co with the data by Michel *et al.* is explained by the difference in the isotopic compositions of the irradiated samples: We used at ITEP enriched $^{56}$Fe samples, while R. Michel *et al.* irradiated samples of natural iron, where $^{57}$Co could be produced via $^{57}$Fe(p,γn)$^{57}$Co and $^{58}$Fe(p,γ2n)$^{57}$Co additional reactions not contributing in our case.

As to the comparison of our results with the inverse kinematics data of Webber *et al.* [42], it is of importance to note that the data of [42] on $^{42}$K and $^{43}$K are overestimated about two-fold compared with both our data and the data obtained elsewhere, the GSI data [43] in particular. It may be admitted, therefore, that the notable differences (up to by a factor of 3) between the data of [42] and [43] presented in Fig. 12 of [43] must be interpreted to be significant systematic overestimations in [42].

The discrepancies in other cases have arisen from different databases used, from different monitor reaction cross sections, and from differences in the methods for determining the absolute and relative spectrometer detection effectivenesses.

The comparison quality deteriorates also because the short-lived nuclides with $T_{1/2} < 4$ hours were not compared, as the relevant data are not presented in other work.

## IV. CALCULATIONS OF CUMULATIVE CROSS SECTIONS FROM INDEPENDENT ONES

As a rule, theoretical models provide the production cross section of each nuclide independently from the possible following decay of nuclide. Just the same cross sections, named usually as independent ones, are measured by the GSI inverse-kinematics method. On the other hand, measurements by the activation method correspond at most cases to the cumulative yields of nuclides produced after a chain of successive β-decays. To compare the data of activation measurements with the GSI data or theoretical model results the cumulative cross sections should be evaluated from the available independent cross sections. The procedure of calculating the cumulative and reduced cumulative yields is described in detail in [49], therefore we recall here only the main ideas. If the radioactive nuclide transformation chain is presented as shown in Fig. 7, then

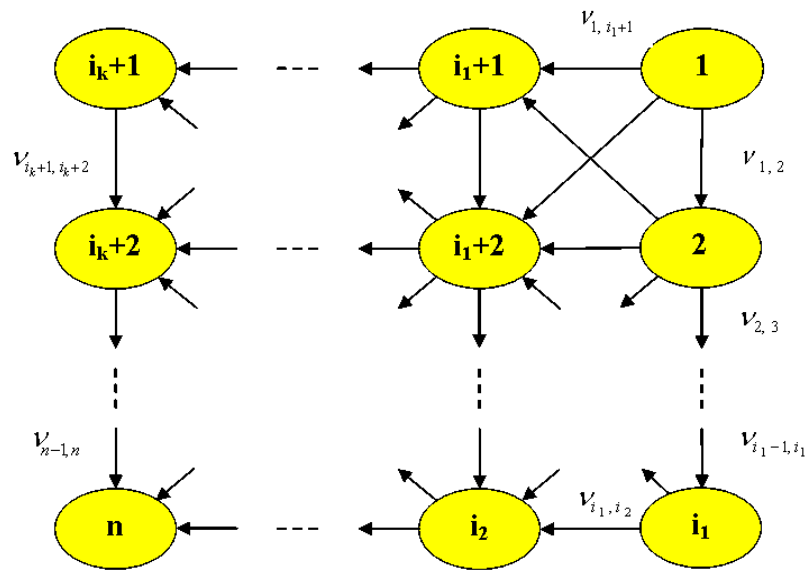

FIG. 7. The radioactive transformation diagram.

the cumulative cross sections can be estimated by the following expression

$$\vec{\sigma}^{cum} = M \cdot \vec{\sigma}^{ind}, \qquad (11)$$

where $M$ is a matrix with elements $m_{kj}$; $\vec{\sigma}^{cum}$ and $\vec{\sigma}^{ind}$ are the vectors, whose elements are the cumulative and independent cross sections of chain elements, respectively.

The matrix elements $m_{kj}$ can be calculated as

$$m_{kj} = \begin{cases} \sum_{i=j}^{k-1} v_{ik} \cdot m_{ik}, & \text{for} \quad k > j, \\ 1, & \text{for} \quad k = j, \\ 0, & \text{for} \quad k < j, \end{cases} \qquad (12)$$

where $v_{ik}$ are the branching factors, which determine the probability of the i-th nuclide to be transformed into the k-th nuclide. The branching factors $v_{ik}$ can be retrieved from the ENSDF database [52] that includes up to 18 modes of radioactive decays: $\beta^-$, $\beta^-n$, $IT$, $\varepsilon$, $\varepsilon+\beta^+$, $p$, $\alpha$, $\beta^+p$, $\beta^+\alpha$, $\beta^+2p$, $\varepsilon p$, $\varepsilon\alpha$, $2\varepsilon$, $n$, $\beta^+$, $2\beta^+$, $2\beta^-$, $2|e$.

## V. THEORETICAL SIMULATIONS WITH VARIOUS CODES

The obtained experimental data were analyzed with 16 different codes, which consider the nuclear reactions at high energies as three-stage processes: the intra-nuclear cascade (INC) of high-energy nucleon collisions followed by the preequilibrium emission of particles with intermediate energies and finally the successive evaporation of low-energy particles from compound nucleus or nuclear fission, if the compound nucleus is heavy enough for fission. The codes used in the present analysis were the LAHET complex [56] with both Bertini [57] and ISABEL [58] options of INC; the MCNPX complex [59] that includes both LAHET options and also two additional code-systems: the CEM2k code with its own models for all three stages [60] and the Liège INCL4 code [61] merged with the GSI evaporation model ABLA [62]. Beside these 6 codes there were used also the upgraded versions of the JINR CASCADE code [63], the Obninsk modification of the LAHET code [64], the Bruyères-le-Châtel BRIEFF complex [65, 66], and the recent LANL complex of the CEM03 and LAQGSM03 codes [67, 68]. A brief discussion of models included in the above codes is presented below.

**A. LAHET**

This code system is widely used during almost three decades for the transport calculations of the high-energy particle interaction with composite targets of a rather complex geometry [56]. It was originally developed as the High Energy Transport Code (HETC) at ORNL [69] and was essentially modified latter at LANL [56]. Due to many new features added at LANL the modified code has been renamed as the Los Alamos High Energy Transport code system (LAHET) and its current versions enable to analyze interaction of nucleons, pions, muons, antinucleons, and light ions with atomic nuclei and any composite materials.

LAHET contains the Bertini INC model [57] to describe nucleon-nucleus interactions below 3.5 GeV and a scaling approximation to extend the energy region to arbitrary high energies, although a reasonable upper limit is about 10 GeV. As an alternative to the Bertini INC, LAHET includes also the ISABEL INC code [58] that is preferable for the incident particle energies below 1 GeV/nucleon.

LAHET calculates preequilibrium processes as an intermediate stage between the intranuclear cascade and the evaporation/fission stage on the basis of the Multistage Preequilibrium Model (MPM) [70], that uses the Monte Carlo method to solve a system of master equations describing the process of equilibration of the excited nucleus remaining after the cascade stage of a reaction. MPM considers the possibility of multiple preequilibrium emissions of n, p, d, t, $^3$He, and $^4$He. Emission probabilities of all these particles are analyzed for the final evaporation stage too.

When the excited nucleus produced after the preequilibrium stage has a mass number lower than 18, LAHET uses the Fermi breakup model [71-74] instead the standard evaporation model to describe the decay of light nuclei. This model treats the deexcitation process as a sequence of simultaneous breakups of the excited nucleus into two or more products, each of which may be a stable or unstable nucleus or a nucleon. Any unstable product nucleus is subject to subsequent breakup again. The probability for a given breakup channel is primarily determined by the available phase space, with probabilities for two-body channels taking into account the Coulomb barriers, the angular moments, and the isospin factors. In the current version of LAHET, only two- and three-body breakup channels are included in spite of a larger number of decay channels considered by the previous breakup calculations on light nuclei [75].

LAHET includes two models for nuclear fission: the Atchison RAL fission-model [76] and the ORNL version [77] based on the Fong fission model [78]. Because iron is a light enough nucleus its fission probability is extremely low and was neglected at all analyzed calculations.

**B. MCNPX**

MCNPX is a next generation of the high-energy transport codes developed at Los Alamos [59, 79]. Its development started in 1994 on purpose to simplify the LAHET applications, to extend the interaction models for high-energy physics, and to provide high-energy calculations with the capabilities of the LANL low-energy transport code MCNP.

Initially, only the LAHET models discussed above were included into MCNPX. However, two alternative models were added to the later MCNPX versions: the CEM2k code based on the Cascade Exciton Model (CEM) developed initially at JINR, Dubna [80] and the Liège INCL4 code by Cugnon *et al.* [61] merged with the GSI evaporation model ABLA [62]. The following features of the added codes should be mention:

i) CEM2k is an improved version of the CEM95 and CEM97 codes described in details in Refs. [81, 82]. Physical models included in the codes are similar as for LAHET, but the concrete realization of INC, preequilibrium, and evaporation models differs in many respects. In particular, the preequilibrium model of CEM takes into account all possible particle-hole transitions, while the LAHET model considers only transitions to more complex particle-hole states (the "never-come-back" approximation). There are also some essential differences between the angular distribution descriptions of emitted particles for the INC and preequilibrium stages. The improvements of CEM2k over the earlier versions relate to a reduction of preequilibrium emission and changes of the interruption conditions for INC and preequilibrium stages.

ii) INCL4 is the latest version of the Liège INC model developed under the HINDAS project [61] and combined all previous model modifications [83-88]. The main improvements of INCL4 relate to an introduction of a smooth nuclear surface, a more consistent implementation of the Pauli blocking principle, and using a much longer stopping time for INC than in previous versions. As result of a long stopping time the intermediate stage of relatively slow intranuclear cascade collisions, which is considered as a preequilibrium one by LAHET, CEM, and many other codes, is described by INCL4 still as the cascade continuation that is followed by the final evaporation stage. Different evaporation codes were tested in a conjunction with INCL4 and the final preference was given to the ABLA code developed at GSI [62].

It should be noted that MCNPX uses more recent and updated values for nuclear masses and binding energies, due to which the MCNPX calculations can differ slightly from the results obtained with the original LAHET, CEM, or INCL4 codes. For similar reasons small differences can exist between the calculations with various MCNPX versions. The calculations presented below were performed with the version MCNPX-2.5.0.

**C. CASCADE-2004**

CASCADE-2004 [63] is an upgrade of the CASCADE a code system [89-92] developed at JINR, Dubna to simulate reactions on both thin and thick targets. It considers nuclear reactions as three-step processes: INC, followed by the preequilibrium stage, followed by evaporation/fission. The INC is described with an old version of the Dubna Cascade Model (DCM) [93], simulating reactions induced by both particles and nuclei. The preequilibrium stage of reactions is described by the Modified Exciton Model (MEM) [94, 95], as realized in the code CEM95 [82]. The evaporation stage of the code was written at JINR (see [90] and references therein) and is based on the Dostrovsky *et al.* model [96]. The fission part of code was also written at Dubna [97] and is based on the Fong fission model [78]. The main difference between CASCADE-2004 and its precursor versions relates to an improved description of the evaporation stage that was developed at JINR by Kumawat [98]. More details on CASCADE-2004 may be found in Refs. [63, 98].

## D. LAHETO

LAHETO [64] is an upgraded and modified version of LAHET used at IPPE, Obninsk. It was developed only for the ISABEL option [58], therefore we do not use it here at incident energies above 1 GeV. In LAHETO, modifications and improvements were made in comparison with LAHET to all three stages of reactions, namely:

i) At the INC stage, corrections were made for nucleon-nucleon and pion-nucleon elementary cross sections to describe better the available experimental data. Actually, the updated elementary cross sections for such interactions from CEM2k [60] were incorporated into LAHETO;

ii) At the preequilibrium stage, improved values for the Coulomb radius parameters were incorporated into the Obninsk version to calculate the charge particle widths;

iii) Finally, the evaporation and fission models of LAHET were modified taking advantage of the IPPE experience on the level density and nuclear-fission analysis (see [99, 100] and references therein). More details on LAHETO may be found in Refs. [49, 64].

## E. BRIEFF1.5.4g

BRIEFF is a code developed at CEA, Bruyères-le-Châtel [65, 66]. It is composed of the INC code BRIC [65], of the evaporation code based on the statistical theory of Weisskopf and Ewing [101], of the modified Fermi breakup model [74], and of the Atchison fission model (the RAL code) [76] slightly modified to be consistent with the evaporation stage.

The INC code BRIC uses a time-dependent approach similar to the Liège INC model. In comparison to the original version [102], the code has been essentially improved for intermediate energies by using the realistic equations of the particle motion inside a nucleus and the in-medium nucleon-nucleon cross sections [65, 103]. These modifications should guarantee that the final steps of INC simulate rather reasonably the preequilibrium stage of nuclear processes. With the same purpose there were estimated more precisely the initialized conditions of hadron-nucleus collisions and the energy distributions of collision products for later steps of the cascade [66].

The essential modifications were also made for the reaction cross sections used for the evaporation stage. Calculations of the partial widths of evaporated particles (p, n, d, t, $^3$He and $^4$He) require the cross sections of a compound-nucleus formation. While in previous versions such cross sections were estimated on the basis of the Glauber model for all charged particles and a combination of the Glauber model with the low-energy optical-model calculations for neutrons, for the current version of BRIEFF 1.5.4g the cross sections of a compound-nucleus formation were calculated directly with the BRIC INC for about 3300 target-nuclei. Such an approach allows removing from the reaction cross sections a contribution related to the preequilibrium processes [66]. For deuterons, tritons and helium-particles the required cross sections were obtained renormalizing the previous reaction cross sections to the ratio of the new and previous ones for protons and their threshold energies taken from the BRIC code. Calculations of the partial widths require also the nuclear level densities that were evaluated on the basis of the back-shifted Fermi-gas model or the energy-dependent level density parameter model by Ignatyuk et al. [99].

The Fermi breakup model replaces the evaporation model when masses of compound nuclei after INC or during the evaporation stage are less than 30. Probabilities of 2- or 3-body decays into light nuclei of mass less than 12 are determined by the available phase-space with an account of limitations on the corresponding Q-values.

**F. CEM03 and LAQGSM03**

CEM03 and LAQGSM03 are the recent versions of the Cascade-Exciton Model and of the Los Alamos Quark-Gluon String Model (LAQGSM) [67, 68]. There are three code versions for each model, which differ by the additional reaction modes included into the codes. The basic versions CEM03.01 and LAQGSM03.01 consider only the traditional stages of nuclear reactions: INC, the preequilibrium processes, and all forms of the successive particle evaporation or fission, as described by a modification of the Generalized Evaporation Model code GEM2 by Furihata [104].

The CEM03.S1 and LAQGSM03.S1 versions include additionally the multifragmentation processes for excited nuclei produced after the preequilibrium stage with the excitation energy above 2A MeV. The Statistical Multifragmentation Model (SMM) by Botvina et al. [105-109] is used to simulate such processes (the extension "S" corresponds to SMM). A total accessible phase space determines the decay probabilities of all possible reaction products in this model, and its detailed description together with a large amount of results obtained for many reactions may be found in Refs. [105-109]. The corresponding SMM code was combined with CEM03 and LAQGSM03 without any modifications.

The CEM03.G1 and LAQGSM03.G1 use the binary-decay GEMINI code [110-114], which realized the asymmetry-fission Moretto model [115] instead of using GEM2 [104]. The extension "G" corresponds to GEMINI in this case. The emission of the lightest particles, from neutron and proton up to beryllium isotopes, is calculated in such an approach on the basis of the standard evaporation model, but the yields of heavier reaction products are simulated by means of the transition-state decay probabilities of asymmetric fission-like configurations. GEMINI is described in details in Refs [110-115] and the corresponding code was combined with CEM03 and LAQGSM03 without modifications.

Improvements of CEM03.01 relative to CEM2k connected mainly with a revision of elementary cross sections, which were updated on the basis of currently available experimental data [116, 117]. New algorithms were developed for a parameterization of the cross sections, for the extrapolation of them to higher energies and for simulation of angular and energy distribution of particles produced in nucleon-nucleon, pion-nucleon, and photon-nucleon collisions.

Some modifications were introduced also in the preequilibrium stage to improve the simulation of complex-particle emission. The coalescence model as described in Refs. [118, 119] was included in CEM03.01 and the corresponding coalesce probabilities were adjusted to the available experimental data on the complex-particle yields for the proton and neutron induced reactions [116]. The Kalbach systematics [120] has been also incorporated to describe angular distributions of both preequilibrium nucleons and complex particles at incident energies up to 210 MeV. For the evaporation stage CEM03.01 uses a modification of the code GEM2 by Furihata [104] that considers evaporation of light nuclei up to $^{28}$Mg simultaneously with possible emission of nucleons and light clusters (d, t, $^3$He and $^4$He) solely considered by most of the evaporation models.

The LAQGSM03.01 code differs from CEM03.01 solely by the INC stage. The INC of LAQGSM03.01 is based on the improved version [121, 122] of the time-dependent intranuclear cascade model developed initially at JINR, Dubna [118]. It uses the experimental elementary cross sections for energies below 4.5 GeV/nucleon and the calculated ones by the Quark-Gluon String Model [123-126] for higher energies to simulate both the angular and energy distributions of INC particles. In contrast to the earlier versions of the Dubna codes [90, 127] and also to CEM03.01 the LAQGSM03.01 code uses a continuous nuclear density distribution, for which there is no a need to consider refraction and reflection of cascade particles inside or on borders of nuclear zones simulated by many other INC codes. It also keeps track on the depletion of the nuclear density during the cascade development (the so-called "trawling effect"). A detailed analysis of nucleon- and pion-induced reactions for targets from C to Am has shown that this effect may be neglected at incident energies below about 5 GeV in the case of heavy targets like actinides and below about 1 GeV for light targets like carbon. However, at higher incident energies the progressive decrease of nuclear density with the development of the intranuclear cascade has a strong influence on the calculated characteristics and the "trawling effect" should to be taken into consideration [90]. Therefore, it was recommend to simulate nuclear reactions with CEM03.01 at incident energies up to about 1 GeV for light nuclei like C and up to about 5 GeV for heavy ones and to switch to a simulation with LAQGSM03.01, that takes into account this effect, at higher energies of transported particles.

## VI. COMPARISON OF EXPERIMENTAL DATA WITH CALCULATIONS

The modeling was carried out at 17 energies from 0.1 to 3.5 GeV to produce smooth excitation functions (EF).

The cumulative yields required for comparison with experimental data were calculated by formula (10). The metastable reaction products were not calculated. The calculation-to-experiment comparison results are presented both qualitatively (as plots) and quantitatively (as mean square deviation factors <F>). The <F> values were calculated by formula

$$<F>=10^{\sqrt{\left\langle\left[\lg\left(\sigma_{cal_i}/\sigma_{\exp_i}\right)\right]^2\right\rangle}} \quad (13)$$

The measured cross sections were simulated by the all codes described in the previous section.

Each of the above presented codes makes use of its own value for the total reaction cross section. To get a correct comparison among the excitation functions obtained by different codes, the calculated results were renormalized to a single reaction cross section value obtained with the Letaw formula [128].

Fig. 8 compares the results of calculating the mass distributions of 300 and 1000 MeV proton-induced reaction products with the experimental data obtained by the inverse kinematics method [43] and with the cumulative yields measured in the present work. Since the cumulative yields correspond to but a fraction of the products, their difference from the GSI data characterizes the contributions from the produced stable isotopes and radioactive isotopes that do not belong to the respective beta-decay chains.

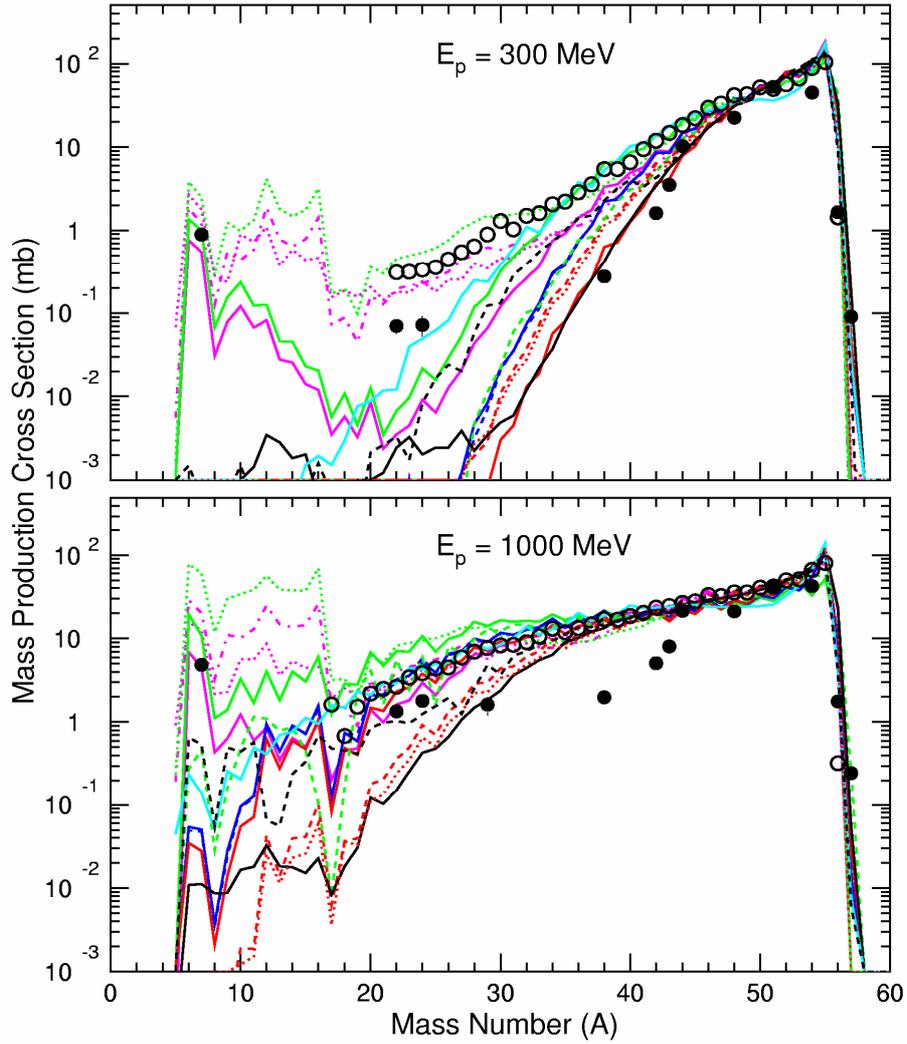

FIG. 8. Mass distributions of $^{56}$Fe(p,x) reaction products measured at ITEP (filled circles) and GSI (open circles) for 300 (top plot) and 1000 MeV (bottom plot) energies as well as simulated by the codes. The codes are designated with the same lines as in Figs. 3-5.

All the models give a sufficiently good description of the mass yields of the products close to target nucleus mass (A>35-40). In the mass range A<30, however, a high-quality description of the observed product nuclide yields is only given by the models that, apart from the conventional evaporation of light nuclei, allow for evaporation of heavy clusters (the CEM and LAQGSM versions). It should be also noted that none of the models gives a good quantitative description of the whole set of experimental data. The CEM03 and LAQGSM models, which are the best to describe the 8<A<18 nuclide yields at 1000 MeV proton energy, predict strong even-odd fluctuations of the yields, which do not seem to affect the experimental data, and give underestimated yields for nuclides with 22<A<32 at 300 MeV proton energy. The experimental data for 1000 MeV proton energy are clearly inclusive of three mass number ranges, namely, A<8, 8<A<20, and A>20. So, our comparison with calculation results produces the impression that different reaction mechanisms dominate in each of the three ranges and, therefore, the qualitative representation of experimental data necessitates a more thorough simulation of each mechanism.

The general regularities of the energy dependence of cumulative yields presented in Figs. 3-5 can be readily accounted for. The reaction products whose mass numbers are only a few unities below the target nucleus (for example, $^{53}$Fe, $^{52}$Mn, and $^{51}$Cr), are produced with sufficiently large cross sections as early as at ~100 MeV, and their yields decrease with increasing the projectile proton energy. This decrease in the yields is due to production of an ever increasing number of reaction products as the energy rises and, because the total reaction cross section varies but little with increasing energy, the yield of a given nuclide decreases as the total cross section gets distributed over an ever rising number of product nuclides.

The reaction products, whose mass numbers differ from target nucleus by more than a dozen of unities ($^{49}$Cr and the lighter nuclides) may be produced only starting from a certain higher threshold energy, which is of the order of the mass difference multiplied by 10-15 MeV. Initially, with increasing the energy of bombarding protons, the yields of such nuclides increase and, after that, begin falling in conformity with the above discussed general regularity of decreasing yields as the number of products increases. The threshold energies of the lightest nuclei are sufficiently high, so only the initial stage of the yield increases with rising the projectile proton energy.

However, the Co isotopes, whose charge is a unity above the target nucleus charge, as well as the $^{56}$Mn isotope, whose yield increases with proton energy (Fig. 3), should be singled out from the above mentioned regularities. At energies below the π-meson production threshold (~140 MeV), the $^{57}$Co isotope is only produced in the reaction of direct or pre-equilibrium proton capture. This reaction mechanism has been rather well studied for projectile neutrons. The reaction cross section is ~1 mb for all medium nuclei and decreases with increasing the projectile energy. Above the π-meson production threshold, the $^{57}$Co isotope can by produced via the (p, π$^0$) reaction, whose cross section increases conforming to the general behavioural regularities of threshold reactions.

The $^{56}$Co isotope is produced in a direct reaction involving excitation of the analogue states of the daughter nucleus. The cross section of such reactions is sufficiently large (comparable with the integral cross section of inelastic proton scattering reactions) and decreases with increasing the projectile energy in conformity with the general regularities of the cross section decrease due to increasing number of reaction products. $^{55}$Co belongs to such products, is produced from $^{56}$Co after neutron emission, and its cross section obeys the general regularities.

Above the π-meson production threshold, the $^{56}$Mn isotope is produced in the $^{56}$Fe(p,pπ$^+$) reaction, whose cross section increases with energy conforming to the high threshold reaction behavioral regularities. Below the meson production threshold, $^{56}$Mn can only be produced in the neutron-induced $^{56}$Fe(n,p) reaction. Such neutrons may be generated due to induced activity in the target chamber. The background effects of this type may be expected to define the difference in the $^{56}$Mn cumulative yields in the given measurements and in inverse kinematics measurements that are free of the neutron background effects.

The general regularities of the energy dependences of reaction product yields discussed above have been corroborated by calculations made by all the models presented in Figs. 3-5. At the same time, the quantitative results are contingent upon a particular model used, so the observed discrepancies follow directly from the differences in the respective models.

TABLE 5. The mean square deviation factors <F> for each of the product group combinations and for each energy based on the ITEP experimental data.

| Code/Model | Product mass (A), Proton energy (MeV) | | | | | | | | | | | | All energies, All products |
|---|---|---|---|---|---|---|---|---|---|---|---|---|---|
| | 300 | | 500 | | 750 | | 1000 | | 1500 | | 2600 | | |
| | A<30 | A>30 | A<30 | A>30 | A<30 | A>30 | A<30 | A>30 | A<30 | A>30 | A<30 | A>30 | |
| MCNPX/INCL | 233 | 5.04 | 141 | 3.19 | 51.5 | 3.09 | 38.1 | 3.08 | 26.1 | 3.30 | 12.1 | 3.47 | **7.36** |

| | | | | | | | | | | | |
|---|---|---|---|---|---|---|---|---|---|---|---|
| MCNPX/CEM2k | -- | 2.73 | 17.2 | 2.49 | 21.1 | 2.57 | 7.83 | 2.72 | 4.87 | 2.88 | 4.02 | 3.15 | **3.64** |
| MCNPX/BERTINI | 1035 | 2.27 | 19.4 | 2.27 | 50.5 | 2.73 | 13.8 | 2.85 | 4.93 | 3.16 | 3.35 | 3.19 | **4.41** |
| MCNPX/ISABEL | -- | 4.04 | 158 | 2.82 | 49.1 | 2.99 | 17.1 | 2.62 | 5.99 | 2.83 | 4.02 | 2.99 | **4.59** |
| LAHET/BERTINI | 542 | 2.29 | 24.9 | 2.26 | 6.98 | 2.66 | 16.5 | 3.15 | 7.34 | 3.37 | 5.69 | 3.14 | **4.09** |
| LAHET/ISABEL | -- | 2.86 | 100 | 2.60 | 44.6 | 3.00 | 15.4 | 3.43 | 7.34 | 3.37 | 5.69 | 3.14 | **4.83** |
| CEM03.01 | 13.0 | 1.81 | 1.99 | 1.88 | 1.32 | 1.88 | 1.49 | 1.92 | 1.58 | 2.04 | 1.72 | 3.17 | **2.24** |
| CEM03.G1 | 2.82 | 2.54 | 2.35 | 2.59 | 2.42 | 2.60 | 2.15 | 2.34 | 1.67 | 2.31 | 1.57 | 3.10 | **2.50** |
| CEM03.S1 | 3.35 | 2.20 | 3.73 | 2.32 | 4.21 | 2.68 | 4.94 | 2.94 | 6.19 | 3.25 | 6.98 | 4.34 | **3.33** |
| LAQGSM03.01 | 45.3 | 2.07 | 6.98 | 1.94 | 3.15 | 2.02 | 2.43 | 2.09 | 1.98 | 2.19 | 1.46 | 3.74 | **2.89** |
| LAQGSM03.G1 | 2.43 | 4.00 | 1.85 | 2.47 | 1.73 | 2.76 | 1.66 | 2.77 | 1.50 | 2.90 | 1.60 | 4.22 | **2.93** |
| LAQGSM03.S1 | 4.64 | 2.79 | 4.35 | 2.41 | 3.75 | 2.67 | 3.89 | 2.67 | 4.17 | 2.66 | 3.59 | 4.13 | **3.10** |
| CASCADE-2004 | 4.69 | 2.70 | 1.87 | 2.84 | 12.4 | 3.13 | 8.00 | 3.72 | 4.55 | 5.43 | 3.04 | 6.48 | **4.27** |
| LAHETO | -- | 4.07 | 108 | 2.43 | 22.8 | 2.83 | 38.9 | 3.24 | -- | -- | -- | -- | **5.45** |
| BRIEFF 1.5.4g | 208 | 2.47 | 12.5 | 3.00 | 8.01 | 3.51 | 6.41 | 3.71 | 5.15 | 3.89 | 3.84 | 3.82 | **4.74** |

To get a better understanding of how the various codes simulate the experimental data, all the reaction products were divided into two groups of the spallation (A>30) and fragmentation (A<30) reaction products, respectively. Table 5 presents the mean square deviation factors <F> for each of the product group combinations and for each energy based on the ITEP experimental data. The energies used in comparisons correspond to the experiment. Table 6 presents analogous results based on the GSI measurements (after convolution of independent yields into cumulative ones using formula (10)). In Tables 5 and 6, the red numerals designate three best codes, while the blue numerals indicate three worst codes. Fig. 9 demonstrates the predictive power of each code (the mean square deviation factors for all energies and all products).

TABLE 6. The mean square deviation factors <F> for each of the product group combinations and for each energy based on the GSI experimental data.

| Программа/Модель | Product mass (A), Proton energy (MeV) | | | | | | | | | | All energies, All products |
|---|---|---|---|---|---|---|---|---|---|---|---|
| | 300 | | 500 | | 750 | | 1000 | | 1500 | | |
| | A<30 | A>30 | A<30 | A>30 | A<30 | A>30 | A<30 | A>30 | A<30 | A>30 | |
| MCNPX/INCL | 153 | 5.85 | 52.0 | 3.30 | 15.8 | 3.10 | 10.2 | 2.25 | 6.63 | 3.27 | **6.06** |
| MCNPX/CEM2k | 1598 | 3.09 | 17.7 | 2.70 | 3.96 | 2.66 | 3.54 | 1.96 | 3.88 | 2.69 | **3.59** |
| MCNPX/BERTINI | 534 | 2.45 | 18.8 | 1.66 | 3.48 | 1.54 | 2.80 | 1.70 | 3.00 | 1.95 | **2.75** |
| MCNPX/ISABEL | -- | 4.76 | 124 | 2.81 | 39.3 | 2.57 | 3.54 | 1.79 | 3.42 | 2.69 | **4.21** |
| LAHET/BERTINI | 1369 | 2.78 | 22.7 | 1.82 | 5.75 | 1.66 | 4.67 | 2.07 | 5.44 | 2.18 | **3.80** |
| LAHET/ISABEL | 1224 | 2.90 | 91.4 | 2.52 | 35.5 | 2.23 | 13.7 | 2.06 | 5.44 | 2.18 | **5.29** |
| CEM03.01 | 27.6 | 1.86 | 2.20 | 2.08 | 1.58 | 2.09 | 1.69 | 1.58 | 1.59 | 2.18 | **2.35** |
| CEM03.G1 | 3.49 | 2.65 | 3.23 | 3.11 | 3.36 | 2.95 | 2.61 | 2.15 | 1.77 | 2.52 | **2.72** |
| CEM03.S1 | 5.42 | 2.68 | 4.91 | 2.62 | 4.07 | 2.70 | 4.83 | 2.51 | 5.50 | 3.28 | **3.10** |
| LAQGSM03.01 | 97.9 | 1.93 | 7.59 | 1.89 | 2.66 | 1.93 | 2.06 | 1.90 | 1.69 | 1.61 | **2.82** |
| LAQGSM03.G1 | 4.22 | 2.26 | 1.97 | 2.64 | 1.77 | 2.85 | 1.53 | 2.76 | 1.48 | 2.46 | **2.52** |
| LAQGSM03.S1 | 13.4 | 3.16 | 5.69 | 2.54 | 3.76 | 2.44 | 4.15 | 2.35 | 4.17 | 2.04 | **2.97** |
| CASCADE-2004 | 4.90 | 2.65 | 1.54 | 2.95 | 1.69 | 2.27 | 1.94 | 2.04 | 1.76 | 2.24 | **2.41** |
| LAHETO | -- | 3.65 | 107 | 2.36 | 20.2 | 2.67 | 34.7 | 2.72 | -- | -- | **5.03** |
| BRIEFF 1.5.4g | 42.2 | 2.05 | 4.86 | 1.93 | 3.18 | 1.83 | 2.99 | 1.77 | 2.62 | 1.72 | **2.58** |

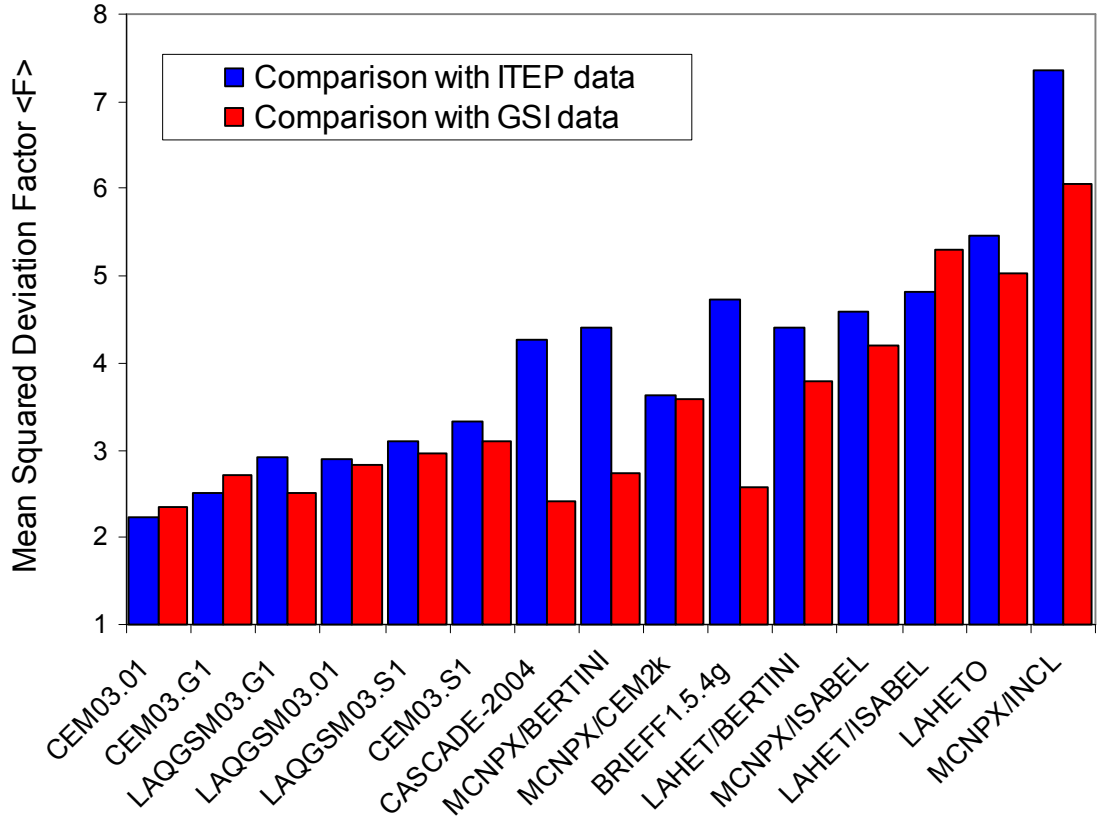

FIG. 9. The predictive power of each code (the mean square deviation factors for all energies and all products).

## VII. CONCLUSION

The work presents results of measuring the production cross sections of radioactive nuclides from 300, 500, 750, 1000, 1500, and 2600 MeV-protons irradiating $^{56}$Fe. In total, 221 independent and cumulative yields of radioactive nuclei of halftimes from 6.6 min to 312 days have been measured. The yields obtained have been compared with results of earlier activation measurements of the cumulative yields of the products of proton interactions with the said Fe isotope and were also compared with the inverse kinematics measurements of the independent product yields from the up-to 1500 MeV proton-induced reactions. Most of the data obtained here are in a good agreement with the inverse kinematics results and disprove the results of some earlier activation measurements that were quite different from the inverse kinematics measurements.

The total experimental data array has been compared with the results by 16 different models. The main discrepancies between different model versions and experimental data have been discussed in brief. The most significant calculation-to-experiment differences are observed in the yields of the A<30 light nuclei, indicating that further improvements in nuclear reaction models are needed, and pointing out as well to a necessity of more complete experimental measurements of such reaction products.

## ACKNOWLEDGEMENTS

We thank Drs. S. Leray, A. Kelić, K.-H. Schmidt, and A. J. Sierk for useful discussions of our results and several valuable suggestions on their presentation in the present paper. This work has been carried out under the EC-supported ISTC Project#3226. The work has also been supported by the Federal Atomic Energy Agency of Russia and, in part, by the U. S. Department of Energy at Los Alamos National Laboratory under Contract DE-AC52-06NA25396.